\documentclass[useAMS,usenatbib]{mn2e} 
\usepackage{epsfig} 
 
\newcommand\aj{{AJ}}%
\newcommand\araa{{ARA\&A}}%
\newcommand\apj{{ApJ}}%
\newcommand\apjl{{ApJ}}%
\newcommand\apjs{{ApJS}}%
\newcommand\aap{{A\&A}}%
\newcommand\mnras{{MNRAS}}%
\newcommand\gca{{Geochim.~Cosmochim.~Acta}}%
   
\newcommand{\eg}{{\rm e.g.\ }}

\newcommand{\ie}{{\rm i.e.\ }}

\newcommand{\cm}{{\rm\thinspace cm}}
\newcommand{\km}{{\rm\thinspace km}}

\newcommand{\s}{{\rm\thinspace s}}
\newcommand{\yr}{{\rm\thinspace yr}}
\newcommand{\erg}{{\rm\thinspace erg}}
\newcommand{\ps}{\hbox{\s$^{-1}\,$}}
\newcommand{\pyr}{\hbox{\yr$^{-1}$}}
\newcommand{\ergps}{\hbox{$\erg\s^{-1}\,$}}

\newcommand{\kmps}{\hbox{$\km\s^{-1}\,$}}
\newcommand{\pcmsq}{\hbox{$\cm^{-2}\,$} }

\newcommand{\halpha}{H$\alpha$}

\newcommand{\hi}{H{\sc i}}
\newcommand{\hii}{H{\sc ii}}

\newcommand{\nH}{\hbox{$N_{\rm H}$}}

\newcommand{\kpc}{{\rm\thinspace kpc}}

\newcommand{\keV}{{\rm\thinspace keV}}

\newcommand{\Lsol}{\hbox{$\thinspace L_{\sun}$}}
\newcommand{\Msol}{\hbox{$\thinspace M_{\sun}$}}

\newlength\dkscolwid
\title[XMM-Newton observations of NGC 6810]{A new superwind galaxy: XMM-Newton observations of NGC 6810}   
\author[D.~K.~Strickland]   
       {David K.~Strickland\thanks{E-mail: dks@pha.jhu.edu} \\
Department of Physics \& Astronomy, The Johns Hopkins University,
3400 North Charles Street, Baltimore, MD 21218, U.S.A.} 
\date{Accepted ..., Received ...; in original ...}     
   
\pagerange{\pageref{firstpage}--\pageref{lastpage}}     
\pubyear{2006}     
     
\begin{document}     
  
\maketitle     
     
\label{firstpage}     
     
\begin{abstract}    

We present the first imaging X-ray observation 
of the highly inclined ($i = 78\degr$) Sab Seyfert 2
galaxy NGC 6810 using {\it XMM-Newton}, which
reveals soft X-ray emission that extends out to a projected height of $\sim 7$
kpc away from the plane of the galaxy.
The soft X-ray emission beyond the optical disk of
the galaxy is most plausibly extra-planar, although it could instead
come from large galactic radius. This extended
X-ray emission is spatially associated with diffuse 
\halpha~emission, in particular with a prominent 5-kpc-long
\halpha~filament on the north-west of the disk. A fraction
$\la 35$\% of the total soft X-ray emission of the galaxy arises
from projected heights $|z| \ge 2$ kpc. 
Within the optical disk of the
galaxy the soft X-ray emission is associated with the star-forming
regions visible in ground-based \halpha~and XMM-Newton Optical Monitor
near-UV imaging.
The temperature, super-Solar $\alpha$-element-to-iron abundance ratio,
soft X-ray/\halpha~correlation, and X-ray to far-IR flux ratio of NGC 6810
are all
consistent with local starbursts with winds, although the
large base radius of the outflow would make NGC 6810 one of the few
``disk-wide'' superwinds currently known.
Hard X-ray emission from NGC 6810 is weak, and the total $E=2$ -- 
10 keV luminosity and spectral shape are consistent with the
expected level of X-ray binary emission 
from the old and young stellar populations. The X-ray observations
provide no evidence of any AGN activity.
We find that the optical, IR and radio properties of NGC 6810 are
all consistent with a starburst galaxy, and that the old
classification of this galaxy as a Seyfert 2 galaxy is probably incorrect.

\end{abstract}    
 
\begin{keywords}    
Galaxies: halos -- Galaxies: individual: NGC 6810 -- Galaxies: Seyfert 
-- Galaxies: starburst --
X-rays: galaxies.
\end{keywords} 
    
\section{Introduction} 
\label{sec:intro} 

NGC 6810 is a early-type spiral galaxy (morphological type Sab(s):sp,
\citealt{rc3}) also classified as a Seyfert 2 \citep[NASA Extragalactic 
Database\footnote{NED: http://nedwww.ipac.caltech.edu/}; 
see also][]{kirhakos90b}. Despite
being relatively nearby ($D\sim 27$ Mpc, see below) it has
not been the target
of much observational study. What little literature
on NGC 6810 exists suggests that a powerful galactic-scale outflow (or
superwind, \citealt{ham90}) emanates from it. \citet{hameed99} present
a \halpha~image of NGC 6810 as part of a study of star formation rates
in early type spirals, but do not comment on the $\sim 5$ kpc-long emission
line filament rising at an angle of $\sim 45\degr$ out of the plane
 of the galaxy
that is visible in their image.
\citet{coccato04} observed NGC 6810 in a survey of ionized gas
along the 
minor axes of spiral galaxies, noting that the high minor axis velocity
dispersion and kinematics were suggestive of an outflow.
The moderately high inclination\footnote{We
  derive an inclination for NGC 6810 of $i=78\degr$ based on the optical
  angular size parameter D25 and R25
  values given in \citet{rc3}. The minor to major axis angular size
  ratio $q = 10^{-R25}$, from which follows the inclination 
  $i = \cos^{-1} \sqrt([q^{2}-q_{0}^{2}]/[1-q_{0}^{2}])$ where $q_{0} = 0.2$
  for type Sab galaxies \citep{haynes84}.}  of NGC 6810, $i = 78\degr$,
is advantageous for studies of extra-planar emission and is similar
to other galaxies with well-studied superwinds such as NGC 253, NGC 3079 and
M82 ($i \sim 76, 78, 85\degr$ respectively, using averages of the
values given in \citealt{strickland04a}).

\begin{figure*}
\psfig{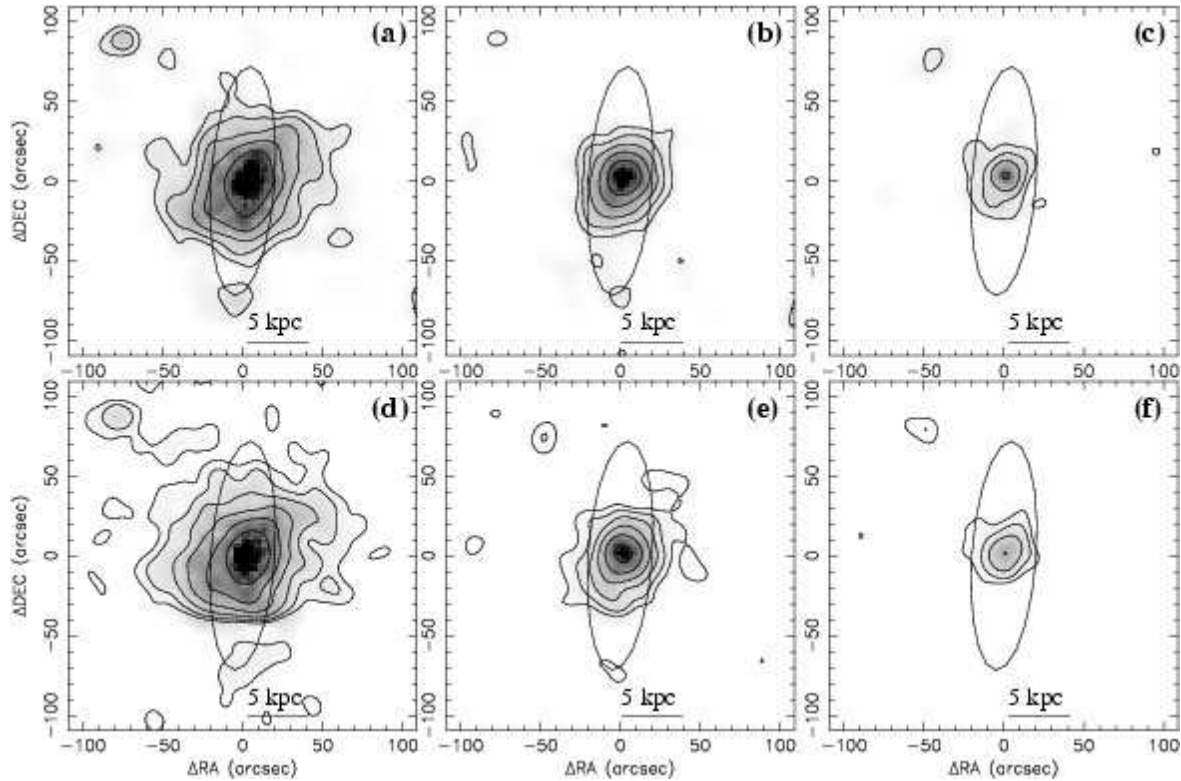}%
\caption{Smoothed XMM-Newton EPIC PN and MOS images of NGC 6810.
Panels a through c are exposure-corrected and background-subtracted
combined MOS1+MOS2 images in the $E=0.3$ -- 1.1, 1.1 -- 2.8 and 2.8 -- 8.0
keV energy bands respectively. The dashed ellipse represents the 
inclination-corrected $d_{25}$ ellipse of NGC 6810 \citep[see][]{rc3}.
The coordinate system shown is the offset from the 2MASS near-IR center of
NGC 6810 at $\alpha=$19:43:34.4, $\delta=$-58:39:20.6 (J2000.0).
Each of the grey-scale images has been 
adaptively smoothed and is shown on a square-root intensity scale.
The overlaid contours are from images smoothed uniformly with a 2-dimensional
Gaussian mask of $FWHM=15\arcsec$. Contours begin at a surface brightness
of $\Sigma = 4.0\times10^{-7}$ MOS 
counts $\ps {\rm arcsec}^{-2}$ and increase in factors of
two. The lowest contours are equivalent to 4.7$\sigma$, 
4.6$\sigma$ and 4.3$\sigma$ in the three energy bands.
Panels d  through f are the equivalent images from the PN detector. Here
the contour levels begin at $\Sigma = 9.6\times10^{-7}$ PN 
counts $\ps {\rm arcsec}^{-2}$, and increase in factors of two. 
The lowest contours are equivalent to 3.0$\sigma$, 
4.7$\sigma$ and 4.1$\sigma$ in the three energy bands.
Extended, 
possibly extra-planar, X-ray emission 
is most clearly detected in the $E=0.3$ -- $1.1$ keV 
soft X-ray band.
}
\label{fig:ximages}
\end{figure*}

Superwinds may play important roles in galaxy formation and evolution
\citep[see \eg][and references therein]{heckman03,veilleux05}, but examples of
starburst-driven winds that are nearby, bright enough 
and have suitable inclinations for detailed study are moderately 
rare\footnote{Within a distance of $30$ Mpc there are approximately
14 galaxies with
inclinations $i \ga 60\degr$ with known or suspected 
superwinds, excluding NGC 6810. 
In order of
increasing distance these are NGC 253, NGC 1482, NGC 1511, NGC 1569, NGC 1808, 
NGC 2146, M82 (NGC 3034), NGC 3044, NGC 3079, NGC 3628, NGC 4631, 
NGC 4666, NGC 4945 and NGC 5775.}.
The existence and properties of AGN-driven superwinds are even less
well constrained \citep{colbert96a,levenson01b,colbert05}.
Given its 
proximity and Seyfert 2 classification we judged that
NGC 6810 deserved investigation.

All currently known superwinds display extended soft diffuse 
X-ray emission from gas at a temperature $\log T \sim 6.5$, 
commonly in close spatial proximity to the warm ionized gas
at $\log T \sim 4$ \citep{dwh98,strickland04a,tullmann06}.
If there is a superwind in NGC 6810 there should be a 5 -- 10 kpc-scale
soft X-ray halo around the galaxy. The presence of the Seyfert nucleus
should also be clear in the hard X-ray energy band, either via direct
AGN X-ray emission or by a reflection component and fluorescent iron line
emission.

\section{Observations and data analysis} 
\label{sec:observations} 

To test these hypotheses NGC 6810 was observed 
with {\it XMM-Newton} for $\sim 49$ ks starting on 
2004 April 25 (Observation ID 0205220101).

We processed the data using version 6.5 of the {\it XMM-Newton}
SAS software, rerunning the EPIC MOS and PN and Optical Monitor (OM)
analysis pipelines
with the latest calibration data. 
{\sc Heasoft} version 5.3.1 and {\sc Ciao} 3.2 
were also used during data reduction.
The recommended standard event grade
filters and proton flare screening was applied to each dataset.
After filtering the remaining exposure times in the XMM observation
were 25.5 ks and 44.1 ks for PN and each MOS instruments respectively.

Spectra and response files were created using SAS and the methods
described in the on-line SAS Data Analysis
Threads\footnote{See {\tt http://xmm.vilspa.esa.es/sas/new/documentation/threads/}}. Spectral fitting was performed using
{\sc Xspec} (version 11.3.1t). 

Simultaneous broad-band optical imaging in V, B, U, UVW1 
($\lambda \sim$ 245 -- 320 nm) and UVM2 ($\lambda \sim$ 205 -- 245 nm) bands
was obtained using the OM in standard imaging mode \citep{mason01}.
These observations achieved
typical limiting magnitudes for a $5\sigma$ detection
of 19.5 (V), 20.7 (B), 20.1 (U), 20.1 (UVW1) and 19.4 (UVM2).
Regions of the images near to NGC 6810 are marred by elliptical 
stray light features caused by internal reflections within the OM telescope.

Copies of the R-band and continuum-subtracted \halpha~images of NGC 6810 
presented in \citet{hameed99} were obtained from the NASA Extragalactic 
Database. 
Astrometric solutions
for these images were calculated based on the V-band OM images, which
match up with Digital Sky Survey images to within $\la 1 \arcsec$.
Pointing offsets during the OM observations were typically also 
$\la 1 \arcsec$.

NGC 6810 is not bright enough an X-ray source to provide useful
X-ray spectra with the grating spectrometers on {\it XMM-Newton} (the RGS).
The RGS data are not discussed further.

\section{X-ray results} 
\label{sec:results} 

\subsection{The spatial distribution of the X-ray emission}
\label{sec:results:xray_spatial}

Images in $E=0.3$ -- 1.1, 1.1 -- 2.8 and 2.8 -- 8.0 keV energy bands
were created from the EPIC PN and MOS data. These energy bands were
chosen based on the X-ray spectrum of NGC 6810 discussed below. The
images from each MOS detector were combined for the purposes of
further image analysis. In each energy band the background intensity 
was estimated as the mean surface brightness on the same
CCD chip as NGC 6810, after excluding the
data within a radius of $30\arcsec$ of any detected point-like X-ray source
or within $90\arcsec$ of NGC 6810. A simple uniform-level 
background  subtraction is appropriate given the relative 
compactness and brightness of the X-ray emission from NGC 6810.

\begin{figure*}
\psfig{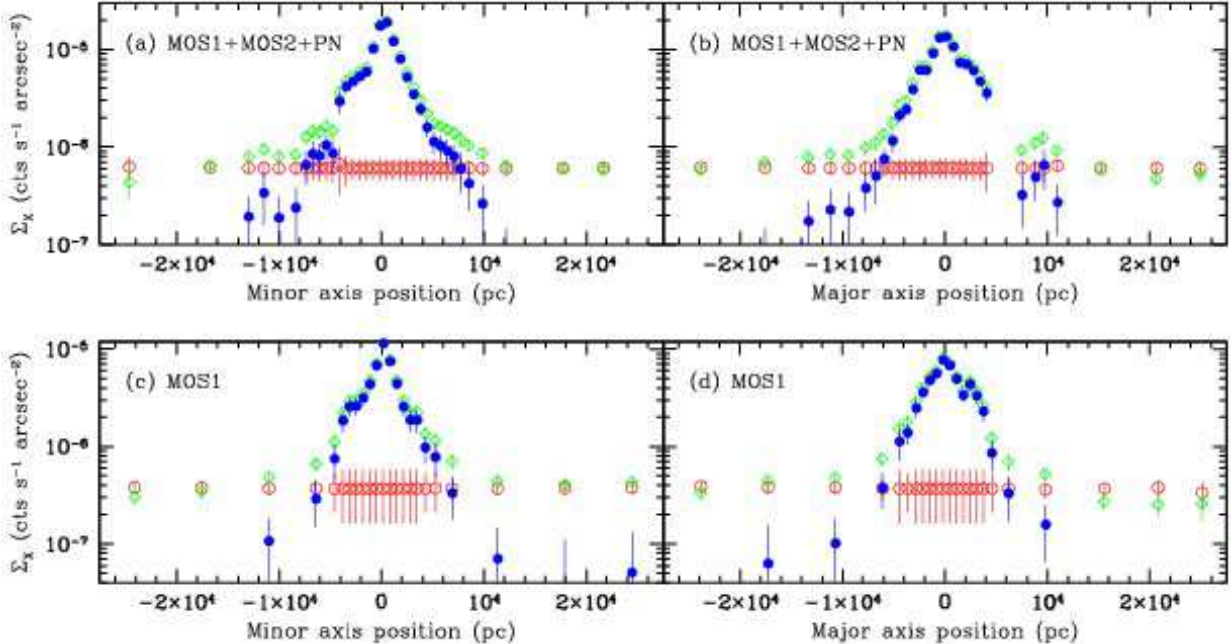}%
  \caption{Soft X-ray ($E=0.3$ -- 1.1 keV) surface brightness profiles 
  along the minor and major axes of NGC 6810. X-ray sources other than NGC 6810
  have been excluded. The total surface brightness is shown as open 
  diamonds (green), with the expected background (from both X-rays and 
  particles) shown as open circles (red). The background-subtracted data
  is shown as filled circles (blue). All of the values were evaluated in
  slices of total depth $76\arcsec$ ($\sim 10$ kpc) perpendicular to the
  axis of interest.
  Panels a and b display the combined MOS and PN data, while panels c and d
  display data from MOS1 alone. Gaps in the data are due to gaps between the
  CCD chips. The error bars represent $1\sigma$ uncertainties. In the minor 
  axis profile the negative axis lies to the east of the disk as seen in 
  Fig.~\ref{fig:ximages}. In the major axis profile the negative axis extends
  toward the north from the center of the galaxy.
  }
  \label{fig:profiles}
\end{figure*}

Figure.~\ref{fig:ximages} shows adaptively smoothed PN and combined MOS
images of NGC 6810 overlaid with surface brightness iso-contours from
images smoothed with a $FWHM=15\arcsec$ Gaussian mask. 
Displaying the PN and combined MOS images separately
is helpful when studying faint extended emission in that it allows us to 
ascertain whether certain features of interest are genuine and which
may be statistical or smoothing artifacts. The lowest contour levels
displayed in these images are between 3 -- 4.7$\sigma$ above the background,
given the $FWHM=15\arcsec$ uniform smoothing
\citep[see][]{hardcastle00}.

In these smoothed images the soft extended X-ray emission can be traced out
to a projected angular separation of $z \sim 60$ -- $70\arcsec$ ($\sim 8$ -- 
9 kpc) from the plane of NGC 6810. The emission
appears to arise from within the optical disk interior to 
a radius of $\sim 50\arcsec$ (6.5 kpc) from the
center of the galaxy. The centroid of the soft X-ray emission is
slightly offset to the north west of the center of the galaxy.
In the hard X-ray band ($E=2.8$ -- 8.0 keV) 
a single point-like 
region of emission, roughly coincident with the center
of NGC 6810, is visible.

Smoothing will artificially increase the apparent extent of the X-ray
emission. To better quantify the projected size of the soft X-ray emission
1-dimensional profiles of the X-ray surface brightness along the major
and minor axes of NGC 6810 were created (see Fig.~\ref{fig:profiles}), 
excluding the emission from X-ray sources not obviously 
associated with NGC 6810.

The profiles display the total surface soft X-ray brightness, the surface
brightness of the expected background (both the genuine X-ray and particle
background) and the background-subtracted emission associated with NGC 6810.
A version of the SAS processing task {\sc edetect\_chain},
altered to correctly treat
the presence of diffuse emission in and around NGC 6810, was used to
detect X-ray sources and create maps of the total background and exposure time
over the entire field of view of each of the MOS1, MOS2 and PN instruments.
The surface brightness profiles were creating using this location-specific
background and exposure time information, excluding statistically 
significant X-ray sources (other than NGC 6810 itself) and detector-specific
chip-gaps, bad CCD pixels and bad 
columns. The values of the MOS and PN backgrounds 
produced by this method is within $\sim 1$\% of the average 
local background values used for 
the images displayed in Fig.~\ref{fig:ximages} and the X-ray spectra
discussed in \S~\ref{sec:results:spectral}. Each bin in the surface brightness
profiles was initially the sum of the data in a region of 
width $2\farcs5$ parallel
to the axis of interest and total length $76\arcsec$ perpendicular to that
axis. Subsequently consecutive bins were rebinned with the aim of achieving
a minimum of 20 counts per final bin. 

In an effort to maximize the signal-to-noise for the faintest emission the
data from the PN and both MOS detectors was combined, and it is these
combined profiles that are shown in the top panels of 
Fig.~\ref{fig:profiles}. In general the surface brightness profiles 
produced from the PN, MOS1 and MOS2 detectors individually 
were very similar to each other (with the PN and MOS2 profiles being the
most similar to each other), 
justifying their combination. The profiles from the
MOS1 detector were the most different, and are shown in the lower
panels of Fig.~\ref{fig:profiles}, but still display the same general
features.

The soft X-ray minor and major axis surface brightness profiles for NGC 6810 
are similar in form to the diffuse emission profiles found in {\it Chandra} 
X-ray observations of highly-inclined starburst galaxies \citep{strickland04a},
although note that in the {\it XMM-Newton} 
data we can not robustly remove X-ray emission from the
point sources in NGC 6810. The surface brightness drops rapidly with 
increasing distance from the nucleus of the galaxy along both minor and
major axes. The minor axis profiles also clearly show the asymmetry in 
the soft extended emission
between the east and west sides of the disk, presumably caused by
absorption within the disk of NGC 6810.

No obvious
 end or edge to the this extended soft X-ray emission can be discerned
from this data. At low signal-to-noise the emission appears to extends out to
projected distances of
$|z|$ or $|r| \sim $10 -- 15 kpc ($\sim 75$ -- $115\arcsec$) 
from the center of the galaxy. However it
is possible that, rather than genuine emission from these locations, we are 
instead seeing emission from the bright central regions of NGC 6810 in the 
very extended, but low surface brightness, wings of the {\it XMM-Newton} PSF.
However, at intermediate distances from the nucleus the extended emission must 
predominantly be genuine, as the central core of the PSF is too compact
(FWHM $\sim 6\arcsec$, Half Energy Width $\sim 12 \arcsec$) to be responsible
for the $\sim 1\arcmin$ diameter region of bright soft X-ray emission
(\eg Fig.~\ref{fig:ximages} or Fig.~\ref{fig:profiles}). 

With no clear edge and the effect of the PSF it is impossible to quantify the
size of the extended soft X-ray emission in a manner that is not 
observation-specific. In common with observations of extra-planar
radio emission from spiral galaxies \citep[\eg][]{dahlem95,dahlem06}
we will quote the extent as the observed size at a fixed statistical 
significance, here evaluated as the size at $S/N = 3$ in the 
background-subtracted surface brightness profiles. Under this definition
the $E=0.3$ -- 1.1 keV emission extends to projected heights of
$z \sim -7.0$ (west) and $z \sim +6.9$ kpc (east) along the minor axis, and
along the major axis to 
projected radii of $r \sim -6.3$ (north) and $r \sim +6.5$ kpc (south).

\subsection{Comparison to UV and optical 
images}
\label{sec:results:xray_optical}

\begin{figure*}
\psfig{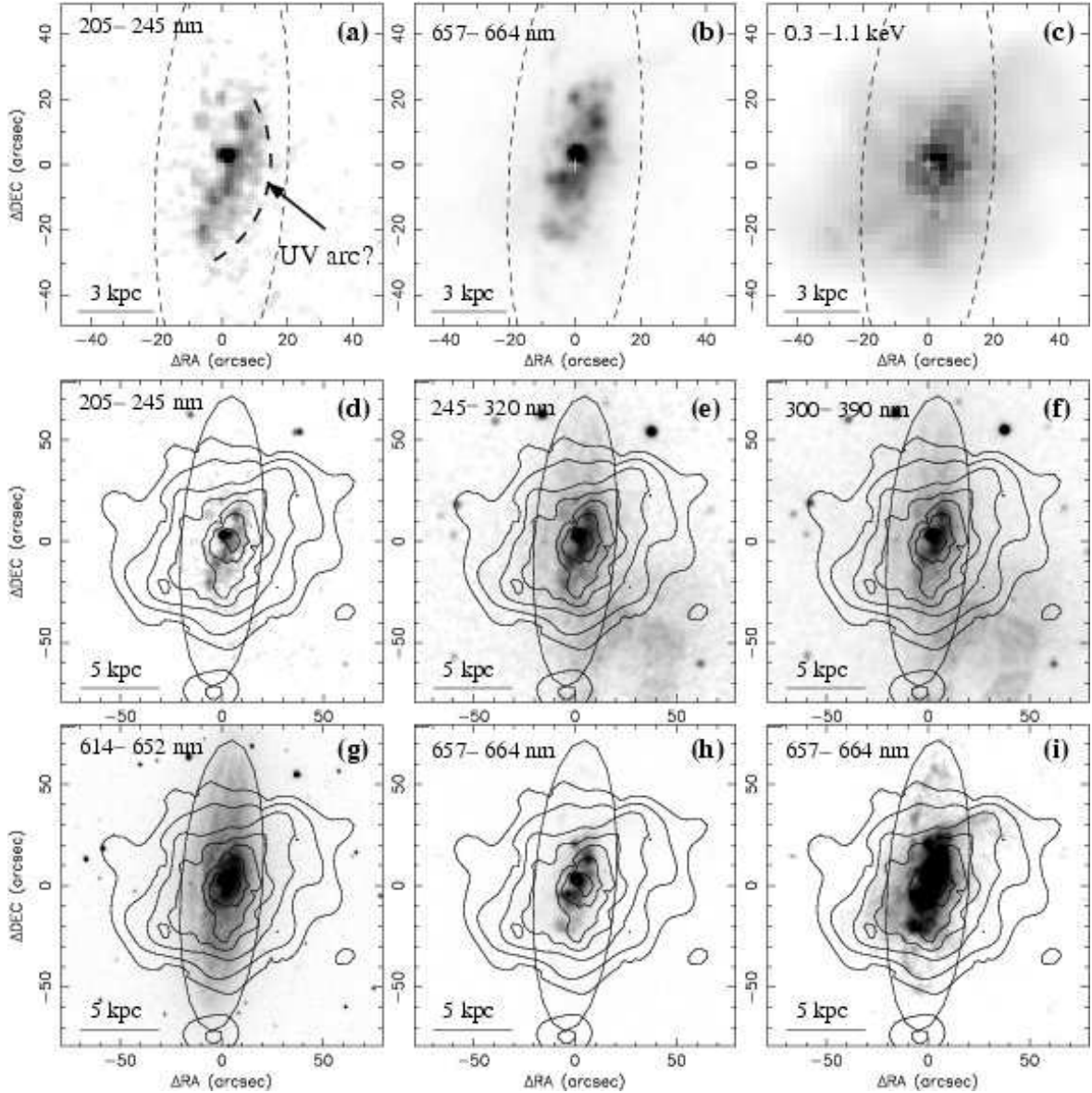}%
\caption{Near-UV and optical images of NGC 6810.  
Panels a to c show the central $90\arcsec$ as seen in the UV 
(UVM2 filter), continuum-subtracted \halpha~and soft X-ray emission.  
The coordinates of the nucleus, as found in the 2MASS survey, 
is shown as a white cross. 
Panels d through f are
{\it XMM-Newton} Optical Monitor images in the UVM2, UVW1 and U filters
respectively, overlaid with the
contours of $E=0.3$ -- 1.1 keV soft X-ray emission from the adaptively
smooth MOS image. Contour levels are at the same surface brightness
levels as in Fig.~\ref{fig:ximages}a. 
Note the stray light artifact to the south west of NGC 6810.
Panels g, h and i are the narrow band R and continuum-subtracted 
\halpha+[{\sc Nii}] images taken by \citet{hameed99}. The \halpha+[{\sc Nii}] 
image is shown in panels h and i on two different intensity scales to 
present both the high and low surface brightness optical emission.
The wavelength
range of each filter is given in units of nm.
The dashed ellipse represents the inclination-corrected $d_{25}$ 
ellipse of NGC 6810.
The strongest spurs of extended soft X-ray
emission seen beyond the optical disk of the galaxy
are clearly associated with \halpha~emission,
in particular at the filament emerging from the northern-most
UV bright region in the UVM2 image and extending $\sim 40\arcsec$ ($\sim 5.3$
kpc)
to the north-west. 
}
\label{fig:optimages}
\end{figure*}

If we compare the X-ray images to ground-based optical and 
near-UV images from the Optical Monitor it becomes apparent
that the soft diffuse X-ray emission is closely associated with
the strongest star forming regions in the disk, and with
warm-ionized gas such as the \halpha~filament that projects beyond
the optical disk of the galaxy (Fig.~\ref{fig:optimages}).

\begin{figure*}
\psfig{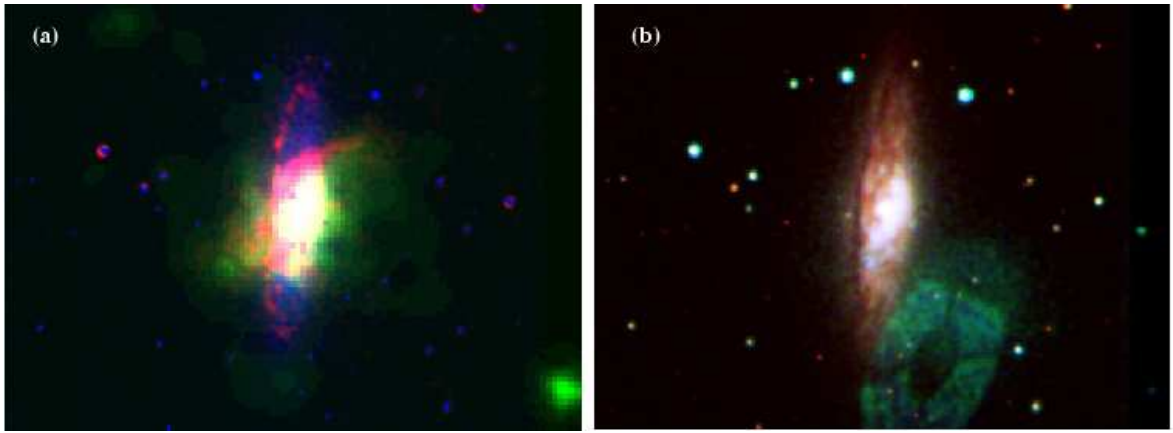}%
\caption{(a) Three-color composite image of NGC 6810, with the
continuum-subtracted \halpha~image (red), adaptively-smoothed
$E=$0.3--1.1 keV MOS1+MOS2 image (green) and R-band image (blue)
all shown on a histogram-equalization scale. The association between
the \halpha~filament on the NW side of the disk and the edge of the
soft X-ray emission is apparent. 
(b) A UV/optical three-color composite image of NGC 6810, shown on the
same spatial scale as (a). The ground-based 
optical R-band image (red), OM U-band (green) and OM UVW1 images (blues)
are all shown on a histogram-equalization scale. Note the prominent dust
lanes in the disk and to the east of the nucleus. The annular structure
to the SW of NGC 6810 seen in the U and UVW1 images 
is an artifact of stray light in the OM telescope. 
}
\label{fig:rgb_images}
\end{figure*}

The most intense soft X-ray emission occurs at the region of brightest
optical and near-UV emission. This region is slightly offset, by 
$\sim 3\farcs5$, to the NNW of
the nuclear position from the 2MASS near-IR observations (2MASS Extended
Objects. Final Release, shown as the white cross in panels a to c of 
Fig.~\ref{fig:optimages}). This offset is larger than the $\la 1 \arcsec$
accuracy of the OM, optical and 2MASS IR image astrometry. 
Strong dust lanes throughout the optical disk of the galaxy
are visible in the 3 color composite image
shown in Fig.~\ref{fig:rgb_images}, and even more so
in an archival F606 HST WFPC2 image which partially covers the nuclear region 
of NGC 6810. Thus the offset of the optical and UV nucleus from the IR
may be due to obscuration. Alternatively the 2MASS position might be
in error, in which case the true nucleus of NGC 6810 lies at 
$\alpha=$19:43:34.3, $\delta=$-58:39:17.3 ($\pm{1\farcs0}$)
(J2000.0).

With respect to the nucleus or the nearby UV/\halpha/X-ray central peak 
the larger-scale UV emission is the most asymmetrically distributed.
What appears to be a UV-bright arc partially cups the nuclear emission
to the west and south, with a total major-axis
diameter of $\sim 30\arcsec$, with 
scattered fainter UV sources at slightly larger radius.
In the \halpha~image the UV arc
and exterior fainter UV sources appear to all be part of a roughly
rectangular \halpha-bright region along Position Angle $\sim 167\degr$,
(offset from the PA of the optical disk by $\sim 9\degr$) and of total
length $\sim 50\arcsec$.
The north
western edge of this region is the base of the most prominent \halpha~filament,
which extends at least $\sim 40\arcsec$ ($\sim 5.3$ kpc) from the plane
of the galaxy (see Fig~\ref{fig:optimages}i).

At low surface brightness the outer edge of the optical disk sports a ring
of \hii~regions coincident with an optical dust lane 
(Fig.~\ref{fig:rgb_images}). This ring is not visible in the soft X-ray
images, even though X-ray emission is detected at regions of similar 
\halpha~surface brightness outside of the optical disk.

The spatial resolution of the EPIC PN and MOS 
instruments is insufficient to allow X-ray emission from compact
sources to be separated from truly diffuse emission.   
The extended soft X-ray emission we observe
is likely to be mainly diffuse thermal emission but this
can not be proven with the existing data. 
Observations with the {\it Chandra} X-ray Observatory 
would certainly allow a better separation of the extended emission into
diffuse and point-like
X-ray sources, in particular in and around the UV and \halpha~bright
star forming regions.

\subsection{Emission region geometry}
\label{sec:results:geom}

\begin{figure*}
\psfig{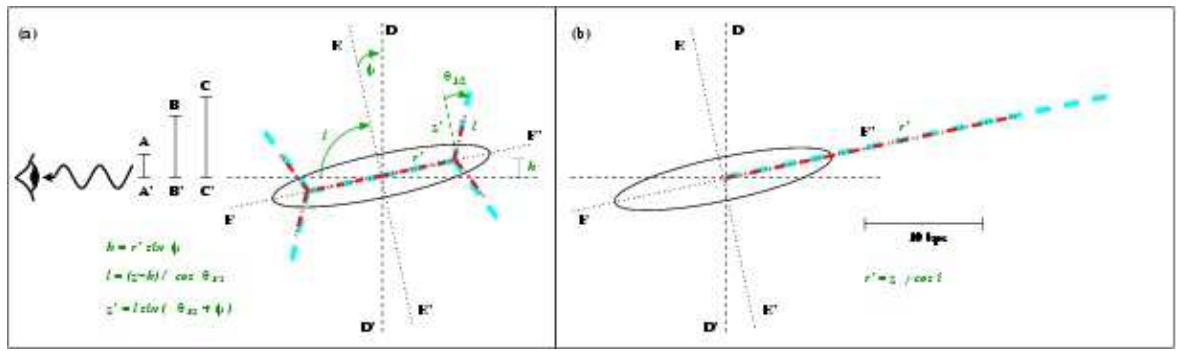}%
\caption{A diagram illustrating two possible distributions for the soft
  X-ray emitting (thick blue dashed lines) and \halpha~emitting material
  (thin red dashed lines) in and around NGC 6810. 
  The panels represent slices perpendicular
  to the plane of the sky, \ie along the line the sight.
    The solid ellipse represents the stellar disk of NGC 6810, with
  a major-axis diameter of 18.9 kpc and axial ratio 5:1 (see below),
  at the adopted inclination of the disk $i = 78\degr$.
  In both panels the plane of the sky is marked by the line DD$^\prime$.
  The true minor axis of NGC 6810 is parallel to the line EE$^{\prime}$,
  and the true major axis is parallel to FF$^{\prime}$.
  The scale bar AA$^\prime$ represents a projected vertical
  height of 2 kpc. The \halpha~filament on the north west side of disk extends
  out to a projected height from the mid-plane of 5 kpc (represented 
  by BB$^{\prime}$), while the soft X-ray emission is detected out to a 
  projected distance of 7 kpc from the mid-plane (CC$^{\prime}$)
  In panel (a) the assumption is that soft X-ray and \halpha~emitting
  plasma are distributed within and above the disk in a geometry similar
  to that of the material observed in galaxies with starburst-driven winds.
  The extra-planar emission is most visible on a truncated 
  conical surface that marks the outer edge of an outflow with a half-opening
  angle of $\theta_{1/2} \sim 25\degr$ (based on the angle between
  the observed \halpha~filament and the projected major axis of the galaxy) 
  and has a radius at its base of
  $r ^{\prime} = 6.5$ kpc. This scenario assumes a cylindrically-symmetric
  geometry.
  In panel (b) it is assumed that the emission lies within the plane of the 
  galaxy and appears in projection above and below the optical disk of the
  galaxy because it extends to much larger galactic radius than the stars.
  To save space no emission is shown on the left hand side of this panel
  although it is assumed to be there.
}
\label{fig:geometry}
\end{figure*}

With an inclination of $i = 78\degr$ the exact location of the 
extended soft X-ray emission is ambiguous with respect to 
the true disk plane of NGC 6810. In particular the soft X-ray emission 
seen in projection beyond the optical disk of the galaxy may indeed 
be extra-planar emission at heights of several kiloparsecs along the 
minor axis, or it
might be emission at large galactic radius but still within the plane
of the galaxy. A scale diagram of these scenarios is shown in
Fig.~\ref{fig:geometry}.

If much of the X-ray and \halpha~emission is associated with a 
starburst-driven wind then it can be seen from Fig.~\ref{fig:geometry}a
that the apparent size of the emission features
(\ie seen in projection) is similar to \emph{but larger than} 
their true vertical extent.
In the example shown the true radius of the base of the outflow is assumed
to be $r^{\prime} = 6.5$ kpc. The true vertical extent of the X-ray emission
$z^{\prime} \sim 4$ kpc can be calculated given that the apparent projected
height $z \sim 7$ kpc is the sum of the projected extent of inclined disk 
($h = r^{\prime} \sin \phi$, where $\phi = 90 - i$) and
the projected size of the edge of the outflow $l \cos \theta_{1/2}$, where
$\theta_{1/2}$ is the half opening angle in the truncated conical
geometry we assume, and that $z^{\prime} = l \sin (\theta_{1/2} + \phi)$.

In this case much of the soft X-ray emission ($E \la 2$ keV) within
the optical confines of the disk and beyond it  
is likely to be genuinely-diffuse thermal emission. Indeed 
fitting the EPIC PN and MOS spectrum of the entire galaxy 
confirms the presence of a soft thermal component (see below).
Note that 
only a minor fraction of the soft X-ray emission comes from regions beyond the
optical disk of the galaxy, and it should be noted that the total
X-ray spectrum presented 
below is not necessarily representative of this (possibly)
extra-planar X-ray emitting material.

If the emission seen at a projected height above the disk mid-plane $z$ is
not actually extra-planar then it must lie at a large galacto-centric
radius, $r^{\prime} = z/\cos(i)$, as shown in Fig.~\ref{fig:geometry}b.
For the optically-derived inclination of $i=78\degr$ for NGC 6810
then the north-western \halpha~filament, with a projected height of 
$z\sim 5$ kpc, would actually extend out to $r^{\prime} \sim 22$ kpc.
The soft X-ray emission extends to a projected height of $z\sim 7$ kpc, 
implying a true radial extent of $r^{\prime} \sim 34$ kpc in this model.
These values are large compared to the stellar radius of the galactic 
disk. Bright \halpha~and/or
soft X-ray emission is not typically observed so far beyond the 
stellar disk of spiral galaxies unless it is extra-planar
\citep{blandhawthorn97,rps97,tyler04}. 

The estimates of the true radial and vertical sizes are sensitive to 
the assumed geometry, in particular 
uncertainties in the inclination of the
galaxy. For example, an 
error of only $5\degr$ in our adopted inclination would change the
radial extent of the X-ray emission in the second case given above
by between 30 and 70\% from the
value estimated above.

As emission from large galactic radius is
less likely than extra-planar emission we will assume 
for the remainder of the paper that
the emission beyond the optical disk of NGC 6810 is extra-planar.

Note that with the existing data we can not prove
unambiguously that this is the case. 
Sensitive high-spatial resolution X-ray observations with
{\it Chandra} might be used to distinguish between the two scenarios.
If the emission geometry is similar to that shown in Fig.~\ref{fig:geometry}a
then we would expect to see evidence of additional absorption of the
X-ray emission near the eastern optical edge of the disk, as the
X-ray emission seen in projection in this region
would need to pass through the intervening disk
of NGC 6810 to reach us (an example of this effect 
can be seen in NGC 253, \eg \citealt{pietsch00}).
We would also need to know more about the \hi~gas distribution
in NGC 6810 before assessing whether such a test were feasible.
\hi~mapping would also be advantageous in providing more accurate
determinations of the inclination angle of the disk.

One explanation for the 
position angle offset between the large-scale optical disk of
NGC 6810 and the UV, \halpha~and soft X-ray bright innermost
6 kpc is that this is a stellar bar. The presence of bars, and
their association with the base of the wind in superwind galaxies,
has been remarked upon before \citep{strickland04a}. However
the existing data is equally consistent with the UV arc and \halpha-bright
region being partially obscured inner spiral arms or a circum-nuclear 
star forming ring.

For the purpose of comparing the presumed extra-planar emission
with other highly inclined ($i \ga 70\degr$) star forming
galaxies observed with {\it Chandra} and {\it XMM-Newton} an estimate
of the fraction of the total (background-subtracted)
X-ray count rate in the $E=0.3$ -- 1.1 keV energy band 
arising at minor axis heights 
$|z| \ge 2 \kpc$ is given in Table~\ref{tab:rates}. The total
count rate was assessed in rectangular region of total width (along the 
major axis of the galaxy)
$2\farcm39$ and total (minor-axis) height $2\farcm54$ ($\sim 20 \kpc$) 
centered on the nucleus. Both MOS and PN data give a similar
answer, with $\sim 35$\% of the soft band X-ray emission coming from
the halo of NGC 6810.
This halo fraction is not as accurate an estimate
of the fraction of diffuse X-ray emission from the halo as might
be obtained with {\it Chandra}, given the larger PSF wings of XMM-Newton,
that we can not separate point source from diffuse emission, and
can not correct for the probably higher absorption experienced by
diffuse emission within the disk of the galaxy. Nevertheless it
is lies within the range of halo diffuse flux fraction
 found for other starburst galaxies with superwinds 
(7 -- 48\%, \citealt{strickland04a}), and supports the hypothesis that
the soft X-ray emission in NGC 6810 extends from the disk into its halo.

\begin{table*}
\caption{X-ray count rates for NGC 6810, accumulated within a
radius of $1\arcmin$ of the center of NGC 6810. The background has 
been subtracted. The halo count rate fraction is the
fraction of the net emission accumulated at
projected minor axis heights of $|z| \ge 2$ kpc.}
\label{tab:rates}
\begin{tabular}{llllll}
\hline
Parameter & Units & \multicolumn{3}{c}{Count rate in each energy band} 
          & Halo count rate fraction \\
          &       & $E=0.5$--2.0 keV & $E=2.0$--10.0 keV &  $E=0.5$--10.0 keV 
          & $E=0.3$--1.1 keV \\
\hline
\hline
PN count rate & cts/s 
              & $0.1193\pm{0.0025}$ 
              &  $0.0106\pm{0.0012}$ &  $0.1299\pm{0.0027}$ 
              & $0.37\pm{0.02}$ \\
MOS1 count rate & cts/s 
              & $0.0313\pm{0.0010}$ 
              &  $0.0039\pm{0.0006}$ &  $0.0352\pm{0.0011}$  
              & $0.35\pm{0.03}$ \\
MOS2 count rate & cts/s 
              & $0.0291\pm{0.0009}$ 
              &  $0.0036\pm{0.0006}$ &  $0.0327\pm{0.0011}$  
              & $0.35\pm{0.03}$ \\
\hline
\end{tabular}
\end{table*}

\subsection{Spectral analysis}
\label{sec:results:spectral}

We created X-ray spectra of NGC 6810 including all emission within
a radius of $1\arcmin$ of the nucleus. No attempt was
made to create multiple spectra of different 
regions within NGC 6810 as the $\sim 100 \arcsec$ 
angular diameter of the X-ray emission 
from NGC 6810 is only of order several PSF widths, and as the
X-ray emission from within the galactic disk is essentially unresolved.

Background spectra were created from a source-free 
circular region of radius $75\arcsec$
centered at $\alpha=$19:43:30.1, $\delta=$-58:37:04.5, 
directly adjacent to the region the
NGC 6810 spectra were extracted from, and also on the same CCD chips.
The background-subtracted count rates for NGC 6810
in each EPIC instrument are given in Table~\ref{tab:rates}.

\begin{figure*}
\psfig{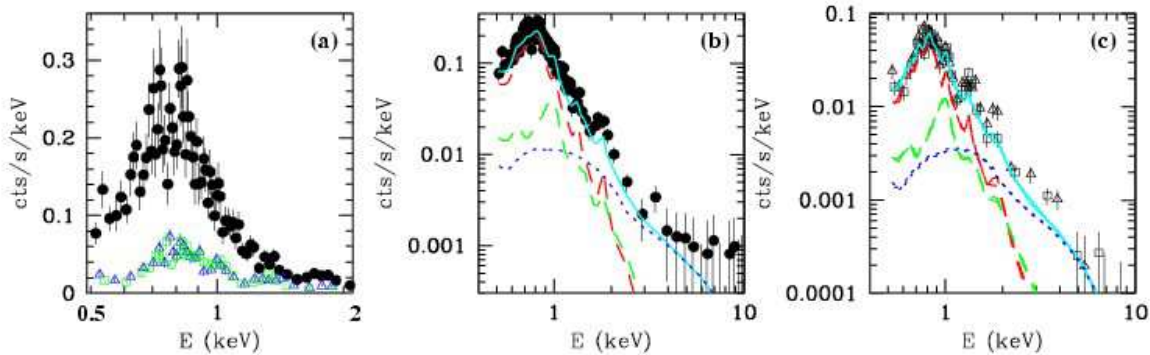}%
\caption{(a) The EPIC PN (filled circles), MOS1 (open triangles) and
MOS2 (open squares) spectra of NGC 6810 shown on a linear flux and 
energy scale, demonstrating the strength of the soft thermal component
at energies below $\sim 1 \keV$ in comparison to higher energy emission.
The PN spectrum is repeated on a logarithmic scale in (b), with solid
lines showing the best-fit 2-temperature thermal plus power law model
fit to the spectrum. The predicted total spectrum is shown as a solid
cyan line, the two APEC thermal plasma components as dash red and green lines
and the power law component as a dotted blue line.
(c) As (b), but showing the MOS1 and MOS2 spectra.
}
\label{fig:spectrum}
\end{figure*}

Spectral models where fit simultaneously to the PN, MOS1 and MOS2 spectra,
allowing for small cross-calibration uncertainties between the
different detectors. The spectra were binned to ensure a minimum of 10 counts
per bin to allow the use of $\chi^{2}$ fitting. As calibration uncertainties
remain significant at low energies we adopt a conservative approach and 
only fit the spectral data in the E=0.5 -- 10.0 keV energy band.

Inspection of the EPIC spectrum of NGC 6810 (Fig.~\ref{fig:spectrum}) reveals
soft thermal emission with line features at energies 
$E \la 2$ keV and a fainter
featureless continuum for $E \ga 2$ keV. This is typical of
CCD spectra of star forming galaxies 
\citep[see for \eg][]{ptak97,dwh98,strickland04a,tullmann06}. There is
no sign of the scattered AGN continuum or strong Fe-K fluorescence that
is commonly found in Seyfert 2 galaxies, including those with starbursts
\citep{levenson01a,levenson02}.

A statistically satisfactory fit was obtained using
a spectral model comprised of an absorbed collisionally ionized 
plasma (we used the APEC plasma code, \citealt{apec01}) to
represent the diffuse hot gas, and an additional more-highly-absorbed
power law to represent X-ray binaries and any low luminosity AGN component.

\begin{table}
\caption{Best fit X-ray spectral parameters for NGC 6810.}
\label{tab:fits}
\begin{tabular}{llll}
\hline
Parameter & Units & \multicolumn{2}{c}{Spectral model} \\
          &       & 1T + PL & 2T + PL \\ \hline
\hline
$\nH_{a}$     & $10^{21} \pcmsq$ 
              & $0.78^{+0.29}_{-0.33}$ & $0.09^{+0.64}_{-0.09}$ \\
$kT_{1}$      & $\keV$
              & $0.57^{+0.02}_{-0.03}$ & $0.54^{+0.03}_{-0.07}$ \\
$Z_{\alpha}$       & $Z_{\alpha, \odot}$ 
              & $0.35^{+0.16}_{-0.10}$ & $0.55^{+0.36}_{-0.23}$ \\
$Z_{\rm Fe}$       & $Z_{\rm Fe, \odot}$ 
              & $0.18^{+0.07}_{-0.04}$ & $0.36^{+0.14}_{-0.15}$ \\
${\rm norm}_{1}$       & $10^{4} \times K$
              & $2.89^{+0.95}_{-0.85}$ & $1.00^{+0.88}_{-0.18}$ \\
$kT_{2}$      & $\keV$
              & \ldots & $1.03^{+0.29}_{-0.17}$ \\
${\rm norm}_{2}$           & $10^{4} \times K$
              & \ldots & $0.31^{+0.22}_{-0.15}$ \\
$\nH_{b}$     & $10^{21} \pcmsq$ 
              & 10 (Fixed) & 2 (Fixed) \\
$\Gamma_{\rm PL}$     & \ldots 
              & $1.73^{+0.74}_{-0.66}$ & $1.90^{+0.58}_{-0.82}$ \\
${\rm norm}_{\rm PL}$     & $10^{4} \times L$
              & $0.15^{+0.17}_{-0.08}$ & $0.17^{+0.13}_{-0.10}$ \\
$\chi^{2}$ (d.o.f) & \ldots
              & 1500.6 (3158) & 1492.6 (3156) \\
$f^{\prime}_{\rm X, tot}$ & $10^{-14} \ergps\pcmsq$ 
              &  $22.0^{+1.2}_{-2.1}$ &  $22.3^{+1.0}_{-3.1}$ \\
$f_{\rm X, tot}$ & $10^{-14} \ergps\pcmsq$ 
              &  $34.3$ &  $25.8$ \\
$f_{\rm X, APEC}$ & $10^{-14} \ergps\pcmsq$ 
              &  $25.0^{+4.9}_{-4.6}$ &  $16.6^{+3.3}_{-3.0}$ \\
$f_{\rm X, PL}$ & $10^{-14} \ergps\pcmsq$ 
              &  $9.3^{+5.5}_{-3.6}$ &  $9.2^{+5.5}_{-3.6}$ \\
$L_{\rm X, tot}$ & $10^{40} \ergps$ 
              &  $3.0$ &  $2.3$ \\
$L_{\rm X, APEC}$ & $10^{40} \ergps$ 
              &  $2.2^{+0.4}_{-0.4}$ &  $1.5^{+0.3}_{-0.3}$ \\
$L_{\rm X, PL}$ & $10^{40} \ergps$ 
              &  $0.8^{+0.5}_{-0.3}$ &  $0.8^{+0.5}_{-0.3}$ \\
\hline
\end{tabular}

\medskip

Note:-- The spectral models used were (in the terminology
  used within xspec) wabs*(vapec+wabs*po) and 
  wabs*(vapec+vapec+wabs*po). Both thermal and power law components
  experience absorption by component $\nH_{a}$, while the power law
  component experiences an additional column  $\nH_{b}$.
  The metal
  abundances (with respect to the Solar abundances of \citet{anders89})
  of both thermal components were assumed to be the same 
  in the two-temperature hot plasma model.  
  The APEC model normalization factor $K = 10^{-14} \times EI / (4 \pi D^2)$ 
  where the emission integral $EI = \int n_{e} n_{H} dV$. The power law
  model normalization factor $L$ is the photon flux in units of  
  photons $\keV^{-1} \ps \pcmsq$ measured at $E=1$ keV.
  Fluxes and luminosities are quoted in the energy band $E=0.3$ -- 8.0 keV.
  The total observed flux is $f^{\prime}_{\rm X, tot}$, \ie no correction
  for absorption has been made. All other fluxes and luminosities were
  absorption-corrected to provide an estimate of the intrinsic flux
  emitted within NGC 6810.
  The confidence regions quoted 
  for fluxes and luminosities are 68.3\% confidence
  for one interesting parameter, all other confidence regions are 90\%
  confidence for one interesting parameter. The error on the observed
  flux $f^{\prime}_{\rm X, tot}$ was calculated using the {\sc Xspec} flux
  Monte Carlo method, all other flux and luminosity errors are more
  approximate as they are based on scaling 
  the uncertainties in the model normalizations.
\end{table}

Using a hot plasma model with Solar abundance ratios \citep{anders89}
results in clear systematics in the spectral fit residuals. The systematics
in the residuals disappeared 
once we fitted for the abundance of the $\alpha$ elements
(specifically the abundances of O, Ne, Mg, Si and Ar were constrained to be
equal) and iron. We obtain an enhanced relative abundance of $\alpha$-elements
to iron (best fit 
$Z_{\alpha}/Z_{\rm Fe} \sim 2 Z_{\alpha, \odot}/Z_{\rm Fe, \odot}$), 
again characteristic of starburst galaxies \citep{strickland04a}.
Note that the absolute values of the X-ray derived elemental 
abundances with respect to hydrogen are ambiguous, given that we
are fitting a simple spectral model to emission most probably coming 
from many spectrally-distinct regions within NGC 6810 
\citep{dahlem00,weaver00} and also because there are 
degeneracies in fitting CCD-spectral resolution X-ray spectra
of $\alpha$-enhanced plasmas \citep{martin02,strickland04a}. 
A two-temperature plasma model
provides a marginally better fit than a single temperature model, but
we would expect the need for a multi-temperature plasma model to 
become more apparent with higher signal-to-noise spectra. The best-fit
spectral parameters for NGC 6810 are given in Table~\ref{tab:fits}.
To ease comparison to other galaxies observed with {\it Chandra} we present
fluxes and luminosities in the $E=0.3$ -- 8.0 keV energy band.
The temperature of the dominant thermal component, $kT \sim 0.6 \keV$,
is typical of values found in fits to the total X-ray emission
from starburst galaxies \citep{rps97,ptak97,dwh98,dahlem03}. We would expect that
there is actually a range of temperatures present, with the extra-planar
emission being on average cooler than the thermal emission from the disk.

The power law component best-fit photon index $\Gamma \approx 1.9$ is again
typical of a purely star forming galaxy. The dominance of the
soft thermal component in the spectrum prevents us from constraining
the absorption column toward the hard X-ray source or sources. We fixed
this column density at $\nH = 2\times10^{21} \pcmsq$, but equally good
spectral fits could be obtained with  $\nH = 1\times10^{22} \pcmsq$.
For the purposes of illustrating the effect this uncertainty has on the
best-fit spectral parameters the best-fit single temperature hot plasma model
plus heavily absorbed power law ($\nH=10^{22} \pcmsq$) 
is also shown in  Table~\ref{tab:fits}.

The similarity of the best-fit power law slope to binary emission
from normal galaxies 
does not alone exclude the presence
of an AGN, as Seyfert 1 galaxies also have hard continuum spectra with 
$\Gamma \sim 1.9$, but the 
hard X-ray luminosity of NGC 6810 ($L_{\rm X} \sim \{4\pm{2}\} 
\times 10^{39} \ergps$, $E=$2--8 keV) is too low for
an unobscured Seyfert galaxy.

\begin{table}
\caption{General properties of NGC 6810.}
\label{tab:n6810}
\begin{tabular}{lll}
\hline
Property          & Value     & Note \\
\hline
\hline
Galaxy type       & Sab(s):sp         & 1 \\
$cz$              & $2031 \kmps$      & 2 \\
Distance $D$      & 27.1 Mpc          & 3 \\
$1\arcsec$ at $D$ & 131.3 pc          &   \\
Major axis P.A.   & 176$\degr$        & 1 \\
$i$                & $78\degr$        & 4  \\
$d^{i}_{25}$       & $2.39\arcmin$, 18.9 kpc   & 4 \\
$v_{\rm rot}$      & $210 \kmps$      & 5 \\
$M_{\rm bary}$     & $8.4 \times 10^{10} \Msol$ & 6 \\
IRAS $S_{12\micron}$ & 1.27 Jy & 7 \\
IRAS $S_{25\micron}$ & 3.55 Jy & 7 \\
IRAS $S_{60\micron}$ & 18.2 Jy & 7 \\
IRAS $S_{100\micron}$ & 32.6 Jy & 7 \\
$S_{60\micron}$/ $f_{100\micron}$ & 0.56 & 7 \\
$\alpha_{\rm IR}$      & 1.87 & 8 \\
$L_{\rm IR}$       & $4.7 \times 10^{10} \Lsol$ & 7 \\
$SFR_{\rm IR}$     & $3.2 \Msol \pyr$ & 9 \\ 
$S_{\rm 4.85 GHz}$ & $72\pm{8}$ mJy & 10 \\
$q = \log (f_{\rm FIR}/f_{\rm 4.85 GHz})$ & 2.57 & 11 \\
$f_{\rm H\alpha}$      & $6.9 \times10^{-12} \ergps \pcmsq$ & 12 \\
$\left[{N~II}\right] 6583 $/\halpha & 0.63, 0.62 & 12, 13 \\
FWHM (\halpha) & $263 \kmps$ & 13 \\
FWHM ($\left[{O~III}\right]$) & $304 \kmps$ & 13 \\
$L_{\rm H\alpha}$      & $6.1\times 10^{41} \ergps$ & 12 \\
$SFR_{\rm H\alpha}$    & $1.88 \Msol \pyr$ & 9 \\
$\log f_{\rm X, APEC}/f_{\rm FIR}$ & -3.77 -- -3.60 & 14 \\
\hline
\end{tabular}

\medskip
\begin{enumerate}{}
\renewcommand{\theenumi}{\arabic{enumi}.}
\item From \citet{rc3}.
\item Recessional velocity from \citet{sanders03}.
\item Assuming $H_{0} = 75 \kmps$ Mpc$^{-1}$.
\item The optically-derived inclination is described in \S~\ref{sec:intro}.
  The inclination-corrected angular
  diameter at 25th magnitude $d_{25}^{i} = 0.1 \times 10^{(D25 - 0.22 
  \times R25)}$ (in arcminutes), where D25 and R25 are given in \citet{rc3}.
\item The inclination-corrected rotational velocity is related to the
  W20 value given in \citet{rc3}: 
  $v_{\rm rot} = 0.5 \times W20 / \sin i$.
\item Baryonic galactic mass $M_{\rm bary} = 10^{9.79} 
  \times (v_{\rm rot}/100 \kmps)^{3.51}$ \citep{bell01}.
\item IRAS flux densities are from \citet{sanders03}, transformed into
  a total 10 -- 1000\micron~luminosity using the equations in
  \citet{sanders96}.
\item As defined in \citet{condon91} the mid-IR spectral slope
  $\alpha_{\rm IR} = \log (S_{60\micron}/S_{25\micron})/\log(60/25)$.
\item Star formation rates derived from the IR or \halpha~luminosity using
  the equations given in 
  \citet{kennicutt98_review}, converted to a Salpeter IMF between
  mass limits of 1 and 100 $\Msol$. The SFR assuming a 
  Kroupa IMF is 1.7 times larger. 
\item Radio flux density at 4.85 GHz from \citet{wright94}.
\item The Far-IR to radio flux ratio $q$ as defined in \citet{condon91}.
\item The extinction-corrected \halpha~flux is based on the 
  \halpha+[{\sc Nii}] flux given in \citet{hameed05}, corrected
  for [{\sc Nii}] emission using the {\sc Nii}/\halpha~flux ratio
  measured by \citet{kewley01}. The \citet{cardelli89} extinction model
  was used to derive A(\halpha) from the E(B-V) value found by
  \citet{kewley01}.
\item From \citet{kirhakos90b}.
\item Logarithm of the total thermal X-ray flux to the FIR flux 
  defined as $f_{\rm FIR} = 1.26 \times 10^{-11} \, (2.58 S_{60\micron} 
  + S_{100 \micron}) \ergps \pcmsq$. 
\end{enumerate}
\end{table}

\section{The nature of NGC 6810} 
\label{sec:discussion} 

The $\sim 7$-kpc-scale extended soft X-ray emission 
discovered in the {\it XMM-Newton}
observations, the previously-discovered 
$\sim 5$ kpc-scale \halpha~filament \citep{hameed99}
and the minor-axis warm ionized gas kinematics \citep{coccato04}
are in combination
convincing evidence for a galactic-scale superwind from NGC 6810.

Can we identify whether the outflow in NGC 6810 is a 
Seyfert 2-or-supernova-driven wind? 
It has already been mentioned that the effective temperature, apparent
$\alpha$-element to iron abundance ratio and halo-to-total X-ray
luminosity ratios of NGC 6810 are consistent with the other local
starbursts with superwinds.
Table~\ref{tab:n6810} is a compilation
of a variety of the pertinent parameters of NGC 6810 derived from
the literature.

NGC 6810 should be classified as a starburst galaxy based on standard 
IR diagnostics using IRAS fluxes. The 60\micron~to 100\micron~flux
ratio of 0.56 is typical of starburst galaxies \citep{lehnert96a,lehnert96b}.
Of course a buried AGN might also elevate the mean dust temperature above
that of a normal spiral galaxy, but we would then expect a lower
mid-IR spectral slope $\alpha_{\rm IR} \propto 
\log S_{60\micron}/S_{25\micron}$. Again NGC 6810 has IR properties typical
of normal starburst, $\alpha_{\rm IR} = 1.87$, significantly higher
than the typical value of $\alpha_{\rm IR} \la 1.2$ of IRAS galaxies
with AGN \citep{condon91}. 

With an total observed 4.85 GHz radio flux of
$72\pm{8}$ mJy \citep{wright94} 
and a radio to FIR flux ratio of $q = 2.57$, NGC 6810 falls
on the radio/FIR correlation for normal and starburst galaxies
(mean $<q> = 2.75\pm{0.14}$, see \citealt{condon91}). 
\citet{forbes98} studied the central few
arcseconds of NGC 6810, finding a low brightness temperature and a q-value
consistent with starburst activity even in this nuclear region.

The star formation rate of NGC 6810 is $SFR \sim 2$ -- $3\Msol \pyr$
based on the galaxy's \halpha~and IR luminosities (these rates
assume a Salpeter IMF between mass limits of 1 and $100 \Msol$).
The core-collapse SN rate associated with this is
approximately one-tenth of the star formation rate, but the associated
kinetic energy input of $L_{\rm SN} \sim (1.2$ -- $2.0) 
\times 10^{42} \ergps$ would be sufficient
to power a superwind (the $L_{\rm SN}$ value comes from the
\citet{leitherer99} starburst model for 
constant star formation rate,
 evaluated 30 Myr after the onset of star formation). 

In local starbursts with superwinds the soft
diffuse X-ray emission associated with the wind typically has a 
luminosity that is only several hundredths of a percent of the IR luminosity of
the galaxy. The {\it XMM-Newton} observations of NGC 6810 lack the spatial
resolution necessary to cleanly separate diffuse from point-like
X-ray emission, but a reasonable approximation is to use the total
estimated soft thermal X-ray luminosity as a proxy for the truly diffuse 
emission. 
The logarithm of the ratio of the thermal X-ray flux to the
FIR flux is $\sim -3.8$ to -3.6 for NGC 6810 (see Tables~\ref{tab:fits} 
and \ref{tab:n6810}). Again these
numbers are consistent with
local starbursts with superwinds, where the log of the soft diffuse
X-ray flux to the FIR flux is $-3.6\pm{0.2}$ \citep{strickland_brazil}.

The soft X-ray luminosity, X-ray/FIR luminosity ratio and IRAS 60\micron~to
100\micron~flux ratio of NGC 6810 are extremely similar to those of
NGC 1511, another starburst galaxy with a probable wind 
observed with {\it XMM-Newton} \citep{dahlem03}. This is despite their
baryonic masses differing by close to an order of magnitude. It illustrates
how strongly correlated the soft X-ray and FIR properties of star forming
galaxies are, and (contrary to na\"ive expectation) 
how little effect large galactic mass appears to have on starburst-driven
superwinds. Galaxy mass does appears to play a role in determining
the critical star formation rate per unit disk area for creation of
radio (and presumably hot gas) halos around spiral galaxies 
\citep{dahlem06}, but for a galaxy of the mass and size of NGC 6810
the transition between galaxies with and without radio halos occurs at
a mean star formation rate per unit area approximately two orders of magnitude
lower than that found in NGC 6810.

The X-ray emission not associated with the thermal components 
(\ie the hard X-ray emission
and the power law component, all presumably from compact objects) 
is also at the level we would expect from
purely stellar processes with no additional AGN contribution required.
\citet{colbert04} found an empirical relationship for 
the X-ray point source luminosity of galaxies from both young and old
stellar populations.
Given the stellar mass of NGC 6810 ($\sim 8 \times 10^{10} \Msol$,
see Table~\ref{tab:n6810}) we would expect the point source X-ray
luminosity from the old stellar population to be $L_{\rm XP, old} \sim 
1.4 \times 10^{40} \ergps$ in the $E=0.3$ -- 8.0 keV energy band. The expected 
X-ray emission from point sources associated with ongoing star formation
is $L_{\rm XP, old} \sim 5.7 \times 10^{39} \ergps$ (we have
corrected for the different IMFs used in the the SF rate given in 
Table~\ref{tab:n6810} and \citet{colbert04}). The total predicted
X-ray point source luminosity of $\sim 2 \times 10^{40} \ergps$ is 
2.5 times larger than the luminosity of the power law component
in NGC 6810, but this is within the level of scatter expected from the 
\citeauthor{colbert04} relationship.

The above mentioned galactic properties and the lack of any sign of
AGN activity in the 0.5 -- 10 keV energy band cast doubt on the classification
of NGC 6810 as a Seyfert 2 galaxy.
The original classification of NGC 6810 as a Seyfert 2 galaxy is due to 
\citet{kirhakos90b}, based on follow-up optical spectroscopy of 
IRAS sources near to or within the HEAO-1 
hard X-ray error boxes of previously unidentified X-ray sources
\citep{wood84,kirhakos90a}. 
\citeauthor{kirhakos90b} identified NGC 6810 as the counterpart to
the
HEAO-1 source 1H1930-589, even though NGC 6810 lies outside the 95\%
confidence error box of this source. This X-ray source had 
a $E=2$ -- 10 keV X-ray flux of $f_{\rm X} \sim (2.0\pm{0.3})
\times 10^{-11} \ergps \pcmsq$, which would correspond to a 
Seyfert-2-like luminosity of $L_{\rm X} \sim (1.8\pm{0.3}) \times 10^{42} 
\ergps$ if the source lies at the distance of NGC 6810.
NGC 6810 also satisfied their optical spectroscopic criteria for classification
as an AGN: [N{\sc ii}]/\halpha~$\ge 1.0$ and/or FWHM ( [O{\sc iii}] or \halpha)
$\ge 300 \kmps$, although just barely as 
only the [O{\sc iii}] line width satisfies their
criterion (see Table~\ref{tab:n6810}). 

A [O{\sc iii}] line width this large would be unusual in a normal galaxy, 
but not in a starburst galaxy with a superwind,
where the velocity of warm ionized gas typically reaches 200 -- 
$600 \kmps$ \citep{ham90,adelberger03}.
Indeed \citet{coccato04} specifically discuss the large line width
in NGC 6810 in terms of a possible outflow. \citet{kewley01} apply
a variety of galaxy classification methods based on optical line
fluxes to IRAS galaxies in the southern hemisphere. 
Of the eight methods they apply to NGC 6810 they classify it
as a H{\sc ii} galaxy six times, with one method giving ambiguous results
and one method yielding a classification of extreme starburst / borderline
LINER. 

\citet{forbes98} also suggest the NGC 6810 may have been misclassified as
being a Seyfert 2 galaxy, based on its starburst-like radio properties
and unpublished optical spectra.  However they
also point out that several Seyfert galaxies with nuclear star
formation have radio and IR properties dominated by the star formation despite
being bona fide Seyfert 2 galaxies.

As we have discussed the hard X-ray spectrum and luminosity of NGC 6810
are inconsistent with a Seyfert 2 classification. The identification
of NGC 6810 with 1H1930-589 is most probably spurious.
That NGC 6810 had a significantly higher X-ray luminosity in the early 
1980's is a less likely alternative, as there is
no other evidence of NGC 6810 ever being a luminous hard X-ray source
(\citet{ward78} place a $3\sigma$ upper limit on the $E=2$ -- 10 keV luminosity
of NGC 6810 of $L_{\rm X} < 6 \times 10^{42} \ergps$ 
based on {\it Ariel 5} data), nor would the 
optical classification of the galaxy
suggest it to be, or have recently been, a Seyfert galaxy.
The possibility remains
that either a low luminosity AGN, or a more luminous but very heavily
obscured AGN, is present in NGC 6810. Deep observations with the Hard
X-ray Detector (HXD) on {\it Suzaku} would be needed to rule the
second possibility out, but at present we conclude that NGC 6810 does
not deserve classification as an active galaxy. 
Radio continuum observations of NGC 6810 would offer another method 
of searching for a low luminosity AGN, in addition to constraining
the location of recent SN activity within the disk.

There is one respect in which the outflow from NGC 6810 may be less than
typical, in that the wind appears to originate not in a compact ($r\la 1$ kpc)
nuclear starburst (as seen in ``classic'' superwind galaxies such as NGC 253, 
M82 and NGC 3079), but to have a large $r \sim 6.5$-kpc radius base.
Such ``disk-wide'' superwinds are less common than the nuclear superwinds,
although there are a handful of previously known examples such as NGC 4666
\citep{dahlem97_n4666} and NGC 5775 \citep{tullmann06}. The base
radius of a superwind appears to match the region of active star-formation
\citep[see \eg][]{strickland_brazil}, and as we have mentioned earlier
may be related to stellar bars (for unknown reasons). Indeed, one
of the greatest current weaknesses of numerical
models of superwinds is their failure to produce a wind base radius equal to
the radius of the star formation region without imposing
unphysical boundary conditions \citep{tenorio-tagle97,ss2000}.

Higher spatial resolution X-ray or optical narrow-band imaging 
of NGC 6810 would
be advantageous in showing whether the extra-planar \halpha~filament and/or
soft X-ray emission physically joins up with
the disk at $r\sim 6.5$ kpc, or whether this is an optical illusion caused
by the projection of a hourglass-shaped nuclear outflow onto the background
disk of the galaxy. \hi~or CO mapping is required to confirm the 
presence of a bar, and would also better constrain the relative geometry
of the hotter ionized gas.

\section{Conclusions}
\label{sec:conclusions}

An observation of NGC 6810 with {\it XMM-Newton} reveals the presence
of extended soft X-ray emission within and beyond the optical disk of the
galaxy. The emission seen above and below the plane of the galaxy might
be emission seen in projection from large galactic radius ($r \sim 30$ kpc),
but is more plausibly extra-planar emission extending to heights of 
$z \sim 7$ kpc from the plane of the galaxy. This, along with 
previously known \halpha~filamentation and peculiar minor-axis
ionized gas kinematics, strongly suggest
 that NGC 6810 is host to a galactic-scale
superwind.

The soft X-ray emission 
seen in projection within the optical disk of the galaxy is closely
associated with star-forming
regions visible in ground-based \halpha~and {\it XMM-Newton} Optical Monitor
near-UV imaging. The actively star forming regions, and apparently
also the base of the hypothesized
outflow, extend out to a radius of $\sim 6.5$ kpc from the nucleus,
and are associated with either a stellar
bar or a partially obscured star-forming ring. The apparently
extra-planar soft X-ray emission is spatially associated with
\halpha~emission, in particular with a prominent 5-kpc-long
\halpha~filament on the north-west of the disk. 

The soft diffuse X-ray emission is presumably thermal bremsstrahlung
and line emission from a collisionally-ionized plasma, 
based on combined EPIC PN
and MOS spectral fitting, and has an intrinsic luminosity of 
$L_{\rm X} \sim (1.5$--$2.2)\times 10^{40}\ergps$ 
in the $E=0.3$ -- 8.0 keV energy band.
The characteristic temperature $kT \sim 0.6$ keV, 
$\alpha$-element-to-iron abundance ratio $Z_{\alpha}/Z_{\rm Fe} \sim 2
Z_{\alpha, \odot}/Z_{\rm Fe, \odot}$, close
soft X-ray/\halpha~spatial association and 
X-ray to far-IR flux ratio $f_{\rm X}/f_{\rm FIR} \sim 10^{-3.6}$ 
are all consistent with local galaxies hosting supernova-driven winds.

The apparently large base radius of the outflow ($\sim 6.5$ kpc) 
makes NGC 6810 one of the few ``disk-wide'' superwinds currently known,
as most local superwinds appear to arise in $\la 1$ kpc-scale
nuclear starburst regions.

Hard X-ray emission from NGC 6810 is weak. The total $E=2$ -- 
10 keV luminosity is $L_{\rm X} \sim (4\pm{2}) \times 10^{39} \ergps$
and the spectrum best fit with a moderately-absorbed 
$\Gamma \sim 1.9$ power law. Based on the stellar mass and star formation
rate of NGC 6810 the observed hard X-ray emission can be fully accounted
for with expected level of X-ray binary emission from
from the old and young stellar population.

The X-ray observations provide no evidence of any AGN activity consistent
with the previous classification of this galaxy as a Seyfert 2.
We find that the X-ray, optical, IR and radio properties of NGC 6810 are
all consistent with a starburst galaxy. The old
classification of this galaxy as a Seyfert 2 galaxy 
is probably incorrect, and can be explained based on the
mis-identification of an unrelated HEAO-1 X-ray source with NGC 6810 and
the broader-than-normal optical emission lines caused by a
supernova-driven outflow.

We conclude that NGC 6810 is an early-type spiral of roughly equivalent
mass to the Milky Way, hosting a superwind with an intriguingly
large base radius.

\section*{Acknowledgments}

It is a pleasure to thank the referee, Michael Dahlem, for the constructive
comments that improved the content of this paper.
This work was funded by grant NNG04G181G from by NASA.
This research has made use of the NASA/IPAC Extragalactic Database 
(NED) which is operated by the Jet Propulsion Laboratory, 
California Institute of Technology, under contract with the 
National Aeronautics and Space Administration.
    


\label{lastpage}    
\end{document}